\documentclass[a4paper]{article}

\usepackage{INTERSPEECH2020}

\title{Transferring Source Style in Non-Parallel Voice Conversion}

\name{Songxiang Liu$^{1,*}$\thanks{*Work done during internship at Tencent AI Lab}, Yuewen Cao$^{1,*}$, Shiyin Kang$^2$, Na Hu$^2$, \\ Xunying Liu$^1$, Dan Su$^2$, Dong Yu$^2$, Helen Meng$^1$}

\address{
 $^1$Human-Computer Communications Laboratory, \\
 The Chinese University of Hong Kong, Hong Kong SAR, China \\
  $^2$Tencent AI Lab, Tencent, Shenzhen, China}

\email{\small \tt \{sxliu,ywcao,xyliu,hmmeng\}@se.cuhk.edu.hk, 
  \small \tt \{shiyinkang,ninahu,dansu,dyu\}@tencent.com}

\begin{document}

\maketitle
\begin{abstract}

Voice conversion (VC) techniques aim to modify speaker identity of an utterance while preserving the underlying linguistic information. Most VC approaches ignore modeling of the speaking style (e.g. emotion and emphasis), which may contain the factors intentionally added by the speaker and should be retained during conversion. This study proposes a sequence-to-sequence based non-parallel VC approach, which has the capability of transferring the speaking style from the source speech to the converted speech by explicitly modeling. Objective evaluation and subjective listening tests show superiority of the proposed VC approach in terms of speech naturalness and speaker similarity of the converted speech. Experiments are also conducted to show the source-style transferability of the proposed approach. 

\end{abstract}

\section{Introduction}

Human speech conveys a wide range of information, among which the linguistic information and speaker identity are the most important. Voice conversion (VC) aims to modify speech characteristics mainly targeting the speaker identity (i.e. voiceprint) of the source speech, while the linguistic information is unchanged. Spectral characteristics related to speaker identity should be modified during the VC process, but other spectral characteristics related to spoken behavior (e.g. emotions, emphasis, etc.) should be retained. This paper will refer to such spoken behavior as speaking style.

Various VC approaches have been proposed and most of them ignore speaking style during conversion. One class of techniques focus only on speaker identity conversion, such as the VC approaches based on Gaussian-mixture model (GMM) \cite{stylianou1998continuous,toda2007voice}, bidirectional LSTM \cite{sun2015voice}, phonetic-posteriorgram (PPG) \cite{sun2016phonetic, liu2018voice}, variational auto-encoder (VAE) \cite{hsu2016voice, hsu2017voice}, generative adversarial network (GAN) \cite{chou2018multi, gao2018voice, kaneko2017parallel, fang2018high, kameoka2018stargan} and etc. These approaches model the source-target speech features frame-by-frame and have limited capability in conducting time-scale modification on the source speech. This may lead to degradation on conversion performance in terms of naturalness of the speech output and its speaker similarity. Another class of techniques uses sequence-to-sequence models and can convert prosodic features such as F0 contour and duration \cite{zhang2019non}, resulting in output that sounds more natural and more similar to the target speaker. However, these approaches inevitably change the speaking style of the source speech during conversion. For example, the source speech may sound excited but the converted speech does not. We consider that voice conversion technologies should preserve such speaking styles and transfer them from the input to the output.

\begin{figure}[t]
  \centering
  \includegraphics[width=6cm]{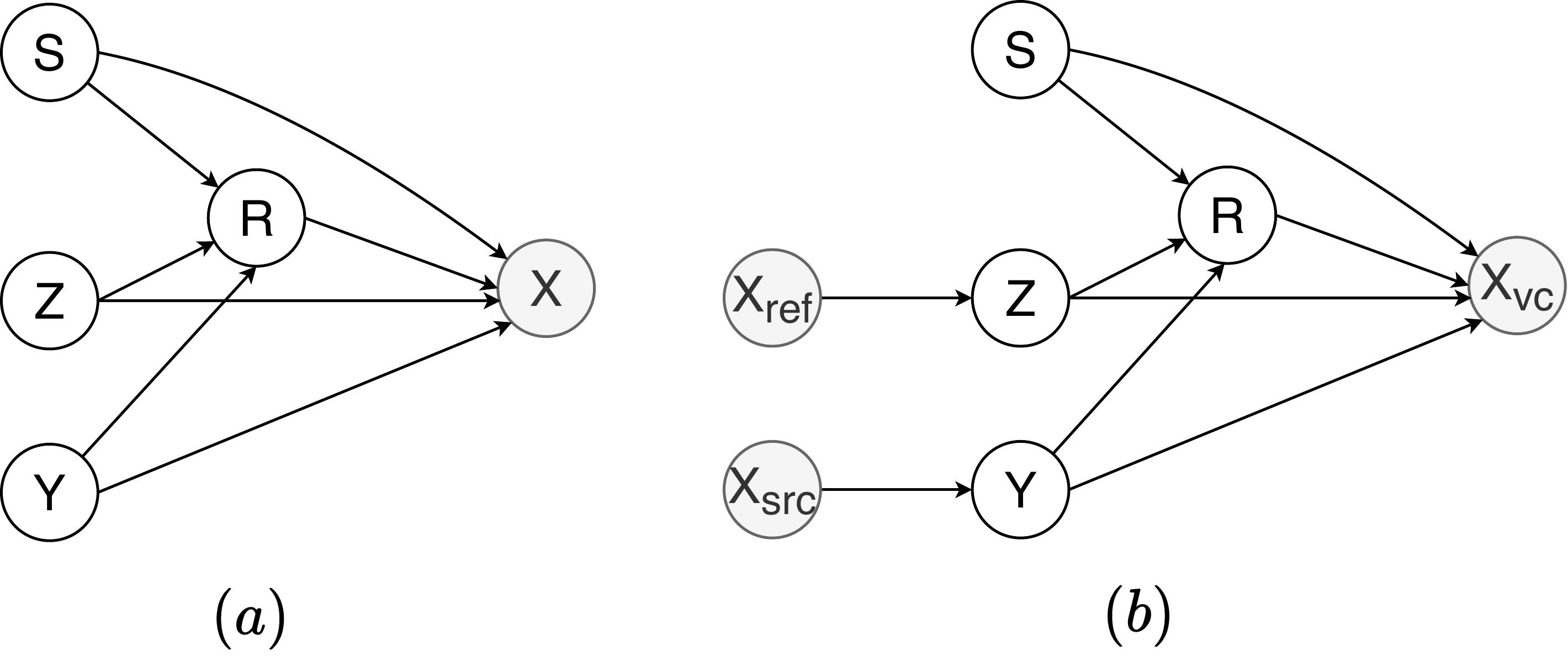}
  \caption{(a) Graphical model for speech generation process. (b) Graphical model for voice conversion process. $X$ and $X_{*}$ represent speech, $S$ represents speaker identity, $Z$ represents speaking style, $Y$ represents linguistic information and $R$ represents rhythm.}
  \label{fig1:speech_production}
\end{figure}

To this end, we propose a sequence-to-sequence based non-parallel VC approach, which has the capability of transferring speaking style from the source speech to the converted speech by explicitly modeling. In this study, we regard the speech generation process as shown in Figure~\ref{fig1:speech_production}(a), where speech $X$ is generated while conditioning on linguistic information $Y$, speaker identity $S$, rhythm $R$ and speaking style $Z$. Rhythm characterizes the duration distribution for realizations of different phonetic structures, which is closely related to the speaking style. We explicitly model rhythm by incorporating a rhythm module into the speech generation model. The conversion process is shown in Figure 1(b). To generate the converted speech $X_{VC}$, the four components ($Y$, $S$, $Z$ and $Y$) are required. $S$ represents speaker identity of a desired target speaker. Linguistic information $Y$ is recognized from the source speech $X_{src}$, while the speaking style $Z$ is inferred from a reference speech $X_{ref}$. When $X_{ref} = X_{src}$, we expect to transfer the speaking style from the source to the converted speech.

The current paper presents an approach which is among the first to explicitly model speaking style throughout the conversion process. The rest of the paper is organized as follows: Section 2 will present the proposed approach. Section 3 will describe baseline approaches for comparison. Experiments and results are presented in Section 4 and Section 5 concludes this paper.

\section{Proposed approach}

The proposed VC system is decomposed into a linguistic content recognizer and a speech generator, which are introduced in the following subsections. Since the recognizer and the generator can be optimized independently, the proposed VC approach does not require parallel speech data between a source speaker and a target speaker.

\subsection{Linguistic content recognizer}

In this study, we use text transcriptions in supervised learning of a linguistic content recognizer. An attention-based sequence-to-sequence automatic speech recognition (ASR) model is used to recognize the phoneme sequence from the speech signal. To speed up training, we incorporate a connectionist-temporal-classification (CTC) \cite{graves2006connectionist} module into the ASR encoder model. The optimization objective during training is:
\begin{equation} \label{eq1}
\begin{split}
\log P(Y|X;\Theta_{rec}) &= \lambda\log P_{CTC}(Y|X;\Theta_{rec}) \\
                         &+ (1-\lambda)\log P_{Att}(Y|X;\Theta_{rec})
\end{split}
\end{equation}
where $P_{CTC}$ is the CTC objective, $P_{Att}$ is the attention decoder objective and $\lambda$ is a hyper-parameter weighting these two terms. Log-Mel-spectrograms are used to represent speech $X$. $\Theta_{rec}$ is model parameters. 

\subsection{Speech generator}

\begin{figure}[t]
  \centering
  \includegraphics[width=4.5cm]{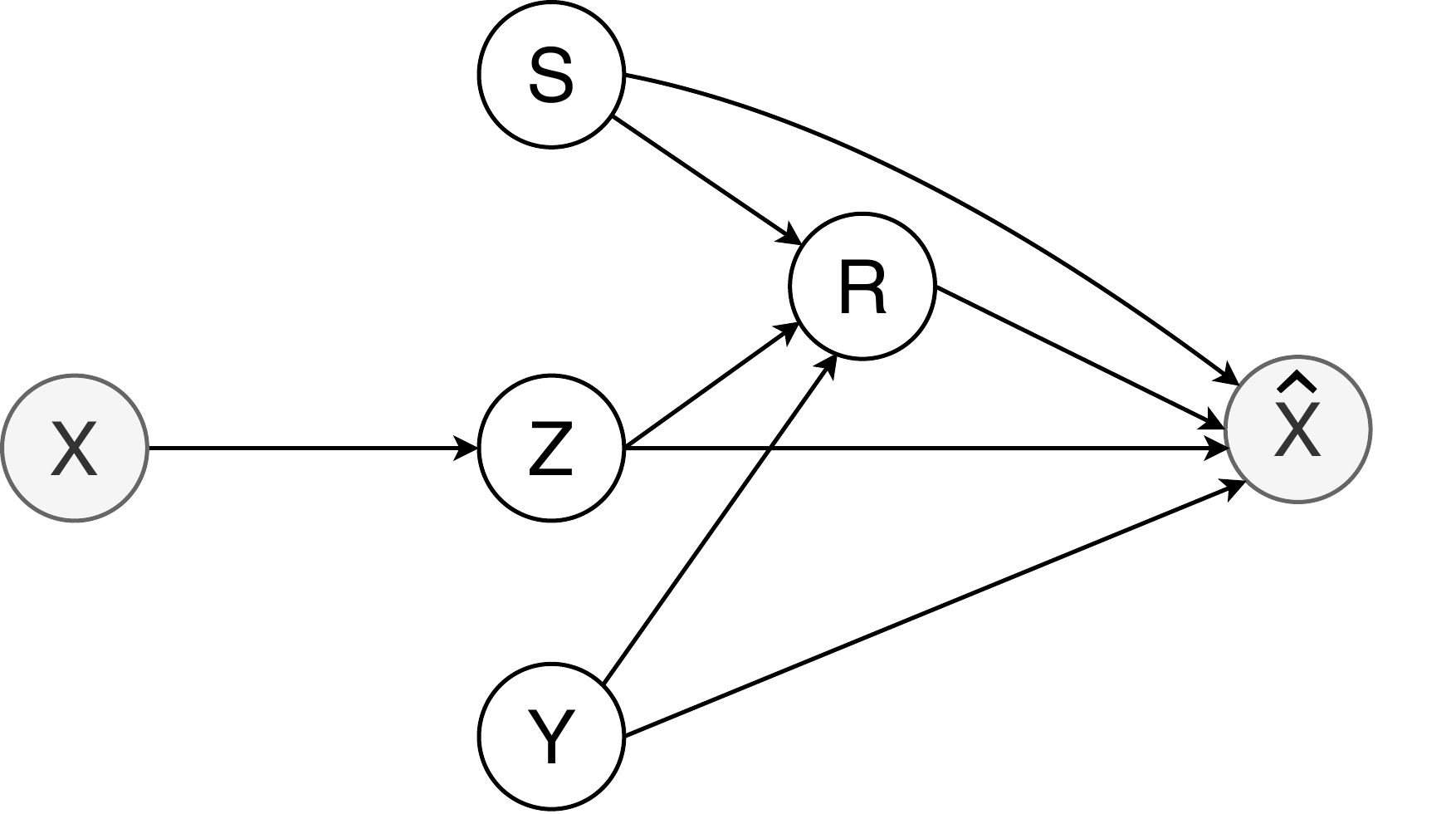}
  \caption{Schematic diagram of speech generator training process.}
  \label{fig2:speech_generator}
\end{figure}

\begin{figure}[t]
  \centering
  \includegraphics[width=8cm]{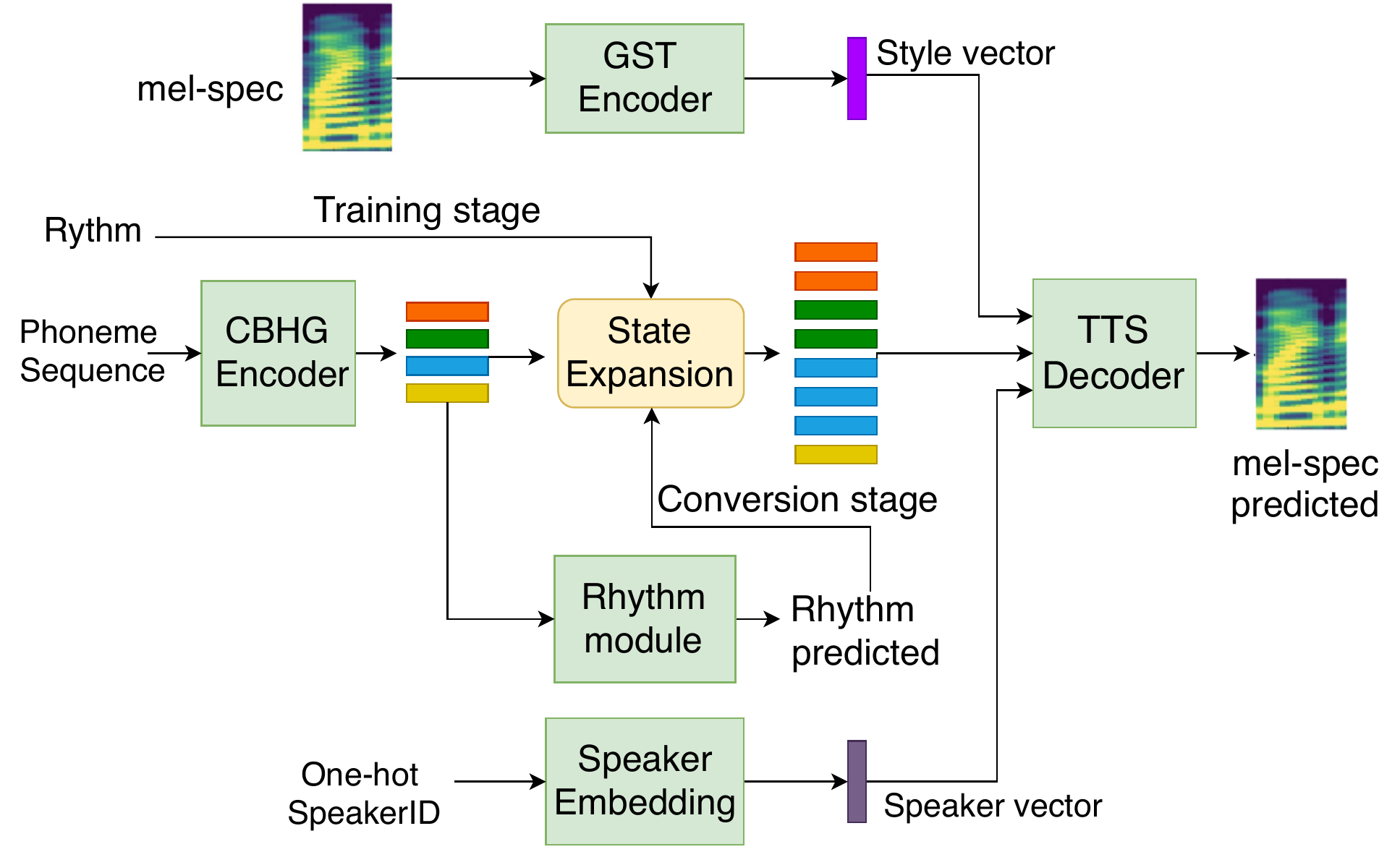}
  \caption{Model architecture of the speech generator.}
  \label{fig3:tts}
\end{figure}

\begin{figure}[t]
  \centering
  \includegraphics[width=6cm]{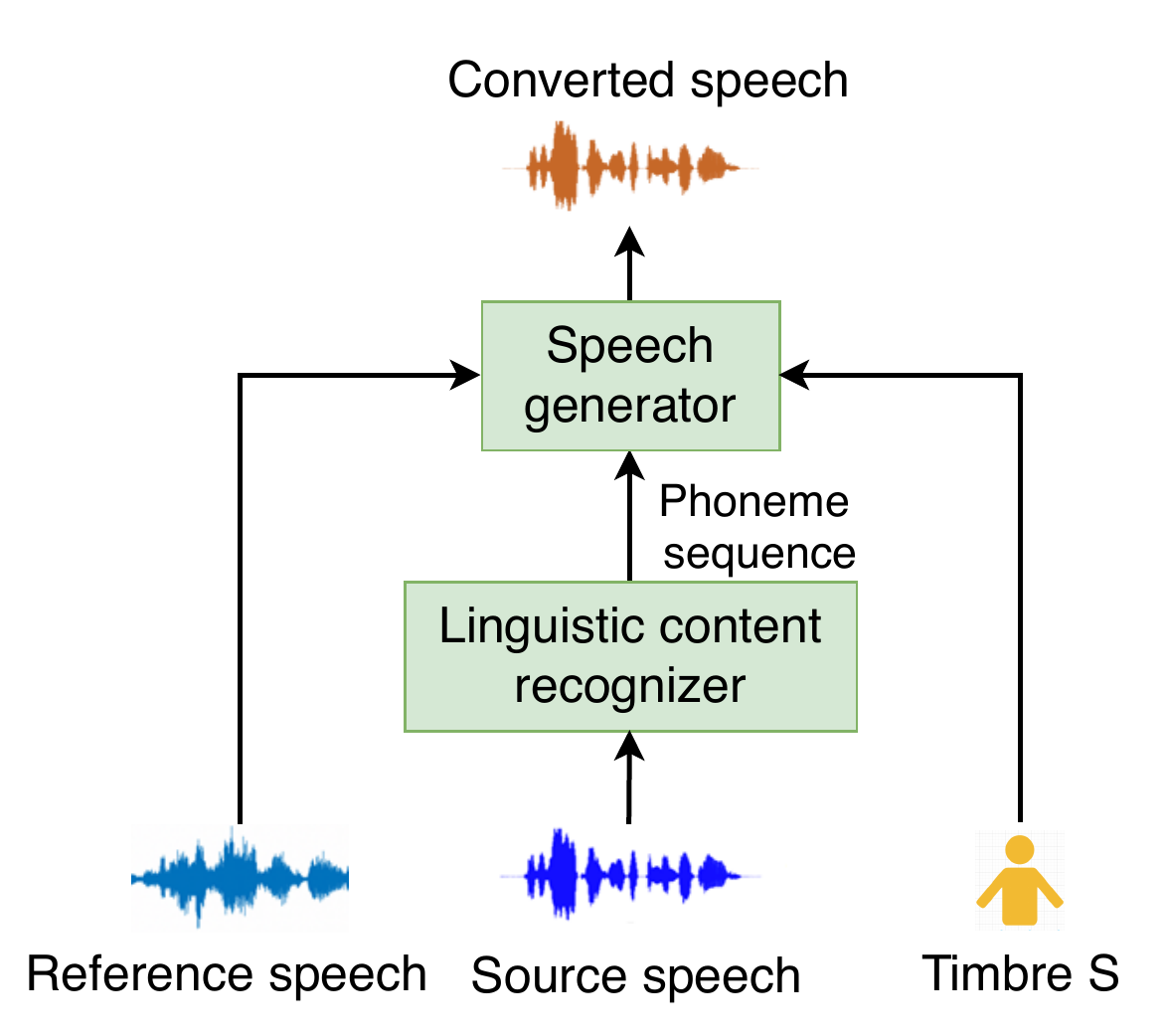}
  \caption{Schematic diagram of conversion process.}
  \label{fig4:coversion}
\end{figure}

We implement the speech generator based on a multi-speaker end-to-end text-to-speech (TTS) model. The training process is shown in Figure~\ref{fig2:speech_generator}, where we use the phoneme sequence as the linguistic representation $Y$ and log-Mel-spectrograms as speech features $X$. We use one-hot representation as the speaker identity input $S$ and learn a speaker embedding table during training. Since the speaking style representation $Z$ is hard to formulate, human annotation is difficult. Hence, for the purpose of our investigation, we infer speaking style from acoustic features $X$ in an unsupervised way. While various techniques have been proposed to learn latent style representations from speech, we follow \cite{wang2018style} and use a global style token (GST) encoder to compute style representation from the acoustic features. We incorporate a rhythm module to predict rhythm $R$ into the speech generator, which functions similarly to the duration predictors studied in \cite{yu2019durian, ren2019fastspeech}. Auto-regressive (AR) TTS models such as Tacotron \cite{wang2017tacotron, shen2018natural} suffer from exposure bias \cite{ranzato2015sequence}, which leads to repeating/skipping words and early stopping phenomenoa in the generated speech. We discover that adding the rhythm module mitigates this issue. There are many ways to represent rhythm, e.g. number of phonemes uttered per second. Here we use an integer sequence for $R$, whose elements correspond to the duration of each phoneme in $Y$. Denoting the rhythm space as $\mathcal{R}$ and speaking style space as $\mathcal{Z}$, during training we want to maximize:
\begin{equation} \label{eq2}
\begin{split}
&P(X|Y, S; \Theta_{gen}) = \sum_{R\sim \mathcal{R}}\int_{Z\sim\mathcal{Z}} P(X, R, Z | Y,S;\Theta_{gen})dZ \\
 & \approx \sum_{R\sim \mathcal{R}}P(X|Y,S,R,Z^*;\Theta_{gen})\cdot P(R|Y,S,Z^*;\Theta_{gen})
\end{split}
\end{equation}
where $\Theta_{gen}$ is the model parameter and $Z^*$ is the computed style representation from $X$ by the GST encoder, i.e., $Z^*=\text{Encoder}_{GST}(X)$.
Since there are infinitely many integer sequences $R\sim\mathcal{R}$, the computation of the summation in Equation~(\ref{eq2}) is intractable. We resort to variational inference, where $P(R|Y,S,Z^*;\Theta_{gen})$ is approximated using a proposed distribution $q(R|X, Y)$, leading to the following evidence lower bound (ELBO):
\begin{equation} \label{eq3}
\begin{split}
\log P(X|Y,S;\Theta_{gen}) &\geq \mathbb{E}_{R\sim q(R|X,Y)}[\log P(X|Y^*,S,Z^*;\Theta_{gen}) \\
&+ \log P(R|Y,S,Z^*;\Theta_{gen})] \\
\end{split}
\end{equation}
where $Y^*$ is the expanded $Y$ along the time axis according to $R$. The first term in the right-hand side of Equation~(\ref{eq3}) is related to the reconstruction loss of the mel-spectrograms while the second term is related to the rhythm prediction loss. While using an AR teacher model to sample $R$ for ELBO optimization in Equation~(\ref{eq3}) is possible as shown in \cite{ren2019fastspeech}, we use an HMM-GMM-based forced-aligner to sample $R$. A LSTM-based AR decoder is adopted to model $\log P(X|Y^*,S,Z^*;\Theta)$ as
\begin{equation} \label{eq4}
\begin{split}
&\log P(X|Y^*,S,Z^*;\Theta_{gen})\\ &= \sum_{t=1}^{R_{sum}}\log P(X_t|X_{<t}, Y^*_{<t},S,Z^*;\Theta_{gen})
\end{split}
\end{equation}
where $R_{sum}$ is the summation of entries in $R$. 

\subsection{Implementation details}
We adopt the transformer-based hybrid CTC-attention ASR model \cite{karita2019comparative} for the linguistic content recognizer.
The input features are 80-dimensional log-mel-spectrograms, on which we conduct utterance-level mean-variance normalization before feeding into the ASR encoder. The Encoder first sub-samples the input features by 4 times in time-scale using two VGG-like max pooling layers. Then the hidden features go through 12 blocks of 8-head multi-head self-attention modules. The ASR decoder consists of 6 blocks of 8-head mutli-head self-attention modules. $\lambda$ in Equation~(\ref{eq1}) is set as 0.3. 

The model architecture of the speech generator is as shown in Figure~\ref{fig3:tts}. The CBHG Encoder has the same structure as in \cite{wang2017tacotron}, which takes phoneme sequence as input. The encoded phoneme sequence is expanded by repetitions along the time axis according to the duration sequence obtained from an HMM-GMM-based forced-aligner. The GST encoder employs the same network structure as in \cite{wang2018style} and produces a 256-dimensional style embedding vector from a mel-spectrogram. One-hot vector is used for speaker identity representation and goes through a speaker embedding layer to get a 256-dimensional speaker vector. The TTS decoder consists of two 128-unit fully-connected (FC) layers, two 512-unit LSTM layers and a post-process module as in \cite{shen2018natural}. The expanded encoder output first goes through the FC layers and then concatenates with the style embedding and speaker vector at every time step. The concatenated features are then fed into the remaining layers of the TTS decoder to obtain the predicted mel-spectrogram.
The rhythm module consists of three 512-unit bidirectional LSTM layers, which is trained to predict rhythm (i.e. duration sequence in this study). What is not shown in Figure~\ref{fig3:tts} for simplicity is that the rhythm module also takes style embedding and speaker vector as additional input, which are concatenated with CBHG encoder output at every frame. We use L2 loss for both mel-spectrogram prediction and rhythm prediction, which correspond to terms 1 and 2 in the right-hand side of Equation~(\ref{eq3}).

\subsection{Conversion process}

The conversion process is shown in Figure~\ref{fig4:coversion}. The linguistic content recognizer first predicts the phoneme sequence from the source speech. Speaking style representations are extracted from reference speech, which can be the source speech if we want to transfer the source style into the converted speech. Speaker identity $S$ is a one-hot vector representing the desired target speaker. Conditioning on the obtained phoneme sequence, speaking style and speaker identity on the speech generator, we get the converted speech, where state expansion operation in Figure~\ref{fig3:tts} uses predicted duration sequence from the rhythm module. In this study, we use a WaveRNN vocoder \cite{kalchbrenner2018efficient} to synthesize waveforms from log-mel-spectrograms. 

\section{Baseline approaches}
We compare the proposed approach with two strong baseline models. \\
\textbf{PPG-VC}: This baseline model is similar to the N10 system \cite{liu2018wavenet} in VCC 2018 \cite{lorenzo2018voice}. The approach consists of a conversion model and a neural vocoder. The conversion model maps PPGs into log-Mel-spectrograms, which consists of four 256-unit bidirectional LSTM layers. We use a WaveRNN model as the neural vocoder. \\
\textbf{Seq2seqVC}: This baseline model is proposed in \cite{zhang2019non} and we use the released implementation by the authors in this study.

\section{Experiments}
\subsection{Experimental setup}
We use an internal multi-speaker Mandarin Chinese corpus for experimental evaluation. The corpus contains 42 speakers (26 female + 16 male). In total, there are 220 hour speech data and the average is 3.5 second per utterance. There are few parallel sentences between any two speakers. We split the corpus into training set (240k utterances), validation set (6457 utterances) and test set (6437 utterances). The sampling rate of the audio is 24 kHz. We first conduct experiments to show that the proposed method has more superior VC performance than the two baseline approaches introduced in Section 3. Then, we show that the proposed approach has better source style transfer performance than the baseline approaches.

\begin{table}[t!]
  \caption{\label{table:1} {ASR-based PER results.}}
  \centering
  \vspace{-0.3cm}
\begin{tabular}{c|cccc}
\hline  
Model             & Sub(\%)     & Del(\%)       &Ins(\%)      &PER(\%) \\
\hline
Ground-truth      &1.5	        &0.1	        &0.1	      &1.7    \\
\hline
PPG-VC            &20.0	        &5.3	        &2.4	      &27.7    \\
Seq2seqVC         &9.6	        &9.5	        &12.6	      &31.7    \\
Proposed          &\bf{4.4}	    &\bf{0.8}	    &\bf{0.4}	  &\bf{5.6} \\
\hline
\end{tabular}
\label{table: mcdrmse}
\end{table}

\subsection{VC evaluation}
In this part, we choose one female (with 37482 training utterances) and one male (with 13863 training utterances) speaker as the target speakers, and choose another female and male speaker as the source speakers.
To evaluate the VC performance of the proposed approach, we use one target utterance uttered in neutral prosodic style as the reference speech in Figure~\ref{fig4:coversion} during conversion. Since VC tasks often assume that there is only limited amount of speech data available from the target speaker, we only use randomly chosen 50 utterances (about 3 minutes) from each target speaker.

During the training process of the proposed approach, we use speech data from all speakers except the two target speakers to train the linguistic content recognizer and speech generator. Then we adapt the speech generator for each target speaker using the 50 utterances from him/her. During the training process of the PPG-VC baseline approach, we first use speech data from all speakers except the two target speakers to pretrain the PPG-to-Mel-spectrogram conversion model, and then we adapt the conversion model for each target speaker using the 50 utterances from that speaker. The training process of the Seq2seqVC baseline approach is similar. We first train the model using speech data from all speakers except the two target speakers. Then, for a given pair of source-target speakers, we use all the training data from the source speaker and the 50 utterances from that target speaker to adapt the model. In this study, we update all model parameters during adaptation. The WaveRNN vocoder is trained using all the training data.

\subsubsection{Objective measure}

We use an off-line trained ASR model to measure the phone error rate (PER) of the converted speech. To show that the proposed approach has more robust conversion performance than the Seq2seqVC baseline, we choose the longest 200 utterances in terms of text lengths from the test set for each source speaker to compute the PER. The results are shown in Table~\ref{table:1},  where Sub, Del, and Ins represent substitution, deletion, and insertion errors, respectively. We can see that the proposed approach achieves lower PER than both baseline models.

\subsubsection{Listening tests}

Two subjective evaluations are conducted: speech naturalness AB test and speaker similarity ABX test. In the AB test, paired speech samples (A and B) from the proposed approach and the baseline approaches are presented to listeners, who are asked to indicate the samples with better naturalness or show no preference (NP). In the ABX test, X indicates the target reference sample. Paired speech samples (A and B) are presented and the listeners are asked to determine which one has closer speaker identity to the reference, or if they are equally close. Each conversion (cross-gender and intra-gender) has 20 samples for evaluation. 10 native Chinese speakers have participated in the evaluations and they are allowed to replay each sample pair as many times as necessary in both evaluations\footnote{Some audio samples in the listening tests can be found in ``https://liusongxiang.github.io/StyleTransferVC/''}.

\begin{figure}[t]
  \centering
  \includegraphics[width=7cm]{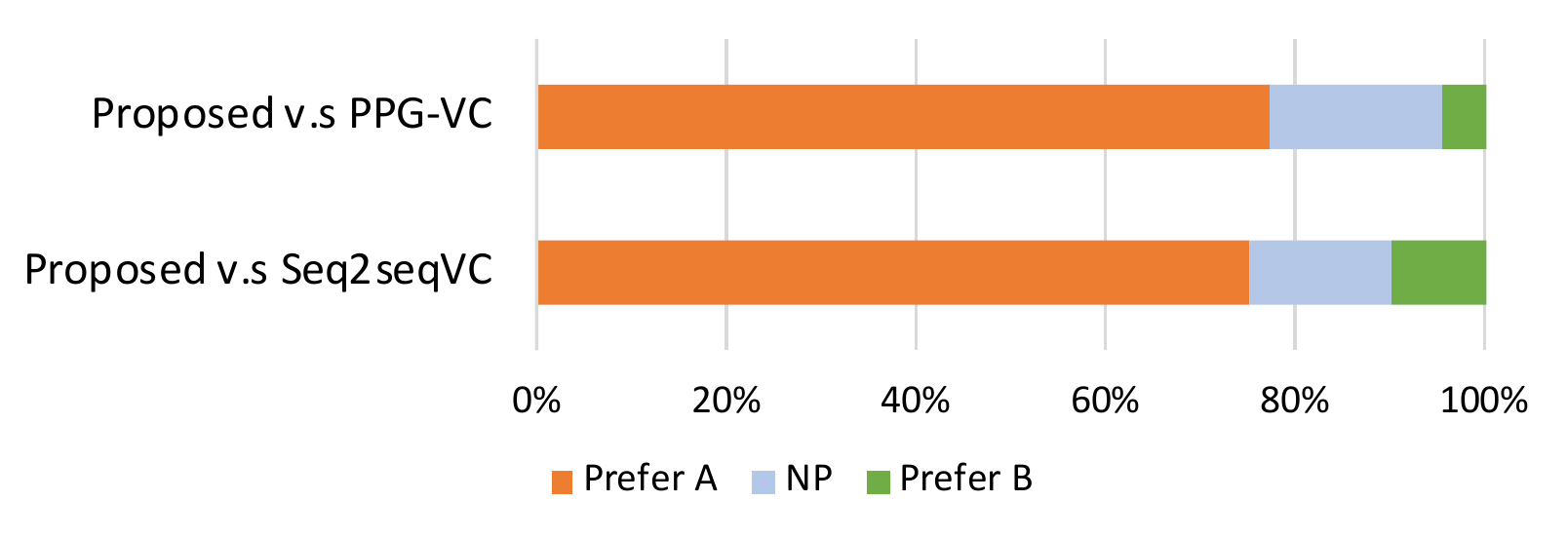}
  \caption{Speech naturalness AB listening test results.}
  \label{fig:naturalness}
\end{figure}

\begin{figure}[t]
  \centering
  \includegraphics[width=7cm]{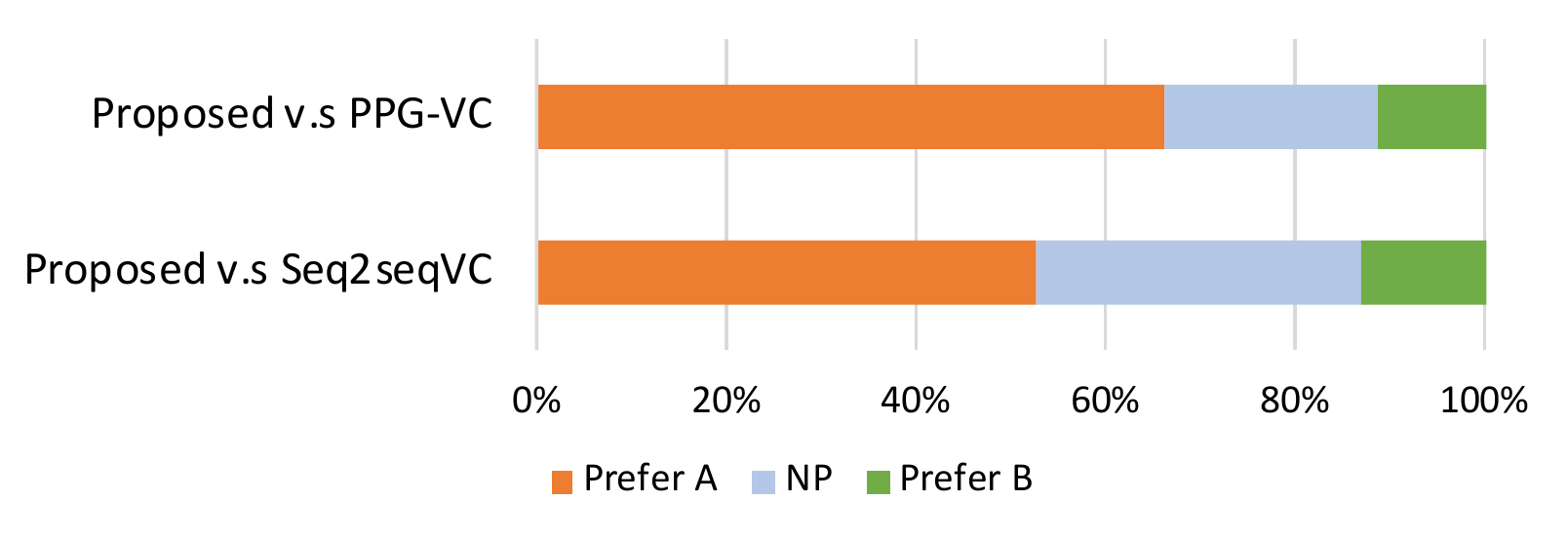}
  \caption{Speaker similarity ABX listening test results.}
  \label{fig:similarity}
\end{figure}

The subjective evaluation results are illustrated in Figure~\ref{fig:naturalness} and Figure~\ref{fig:similarity}. We can see that the proposed approach significantly outperforms the baseline approaches in terms of speech naturalness and speaker similarity of the converted speech. 

\subsection{Style transfer evaluation}

In this part, we evaluate speaking style transfer performance from the source speech of the proposed approach. We choose a female as source speaker (with 7514 training utterances) and a female as target speaker. The training process is similar to that in Section 4.2, except that during adaptation, we use all the training data (6725 utterances) from the target speaker. During conversion, we make the reference speech in Figure~\ref{fig4:coversion} the same as the source speech for source speaking style transfer.

\subsubsection{Visualization}

\begin{figure}[t]
  \centering
  \includegraphics[width=8cm]{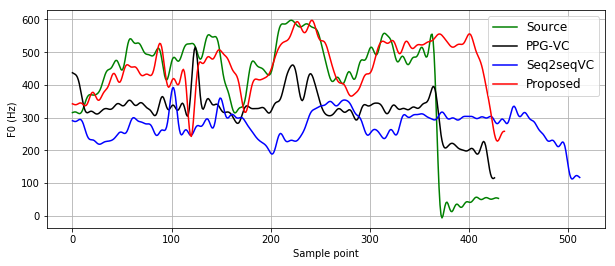}
  \caption{Continuously interpolated F0 contour.}
  \label{fig:f0}
\end{figure}

The F0 contour reflects prosodic variations in an utterance, which is related to speaking style.
Continuously interpolated F0 contours of the converted speech of a source sample utterance are shown in Figure~\ref{fig:f0}. We can see that the F0 contour of the converted speech by the proposed approach matches the source F0 contour more closely than the baseline approaches. Moreover, the converted speech by the proposed approach has greater durational similarity in the source speech than the Seq2seqVC approach. Therefore, we state that the proposed approach performs better in preserving the speaking style of the source speech. The PPG-VC approach employs frame-to-frame framework for the conversion model. So the converted speech by the PPG-VC approach has exactly the same duration as the source speech. However, since the PPG-VC approach adopts a simple linear transformation in logarithmic scale on the source F0, the style transfer performance should be worsened.

\begin{figure}[t]
  \centering
  \includegraphics[width=7cm]{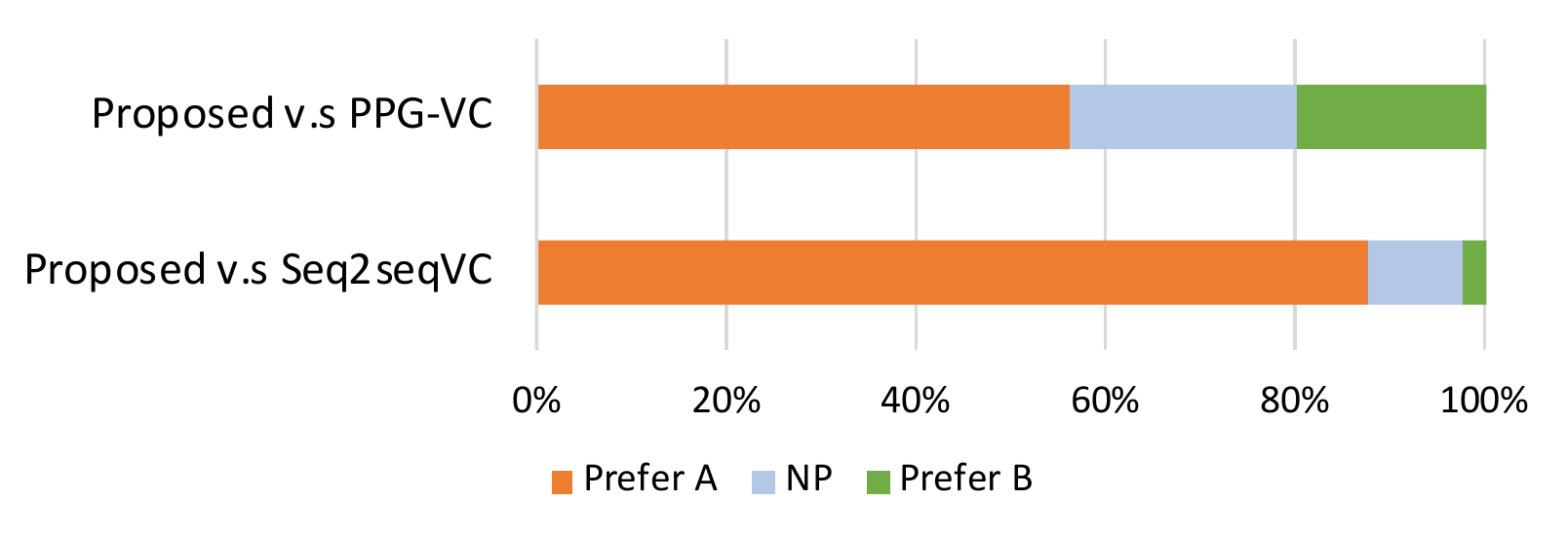}
  \caption{Source style transfer ABX listening test results.}
  \label{fig:style}
\end{figure}

\subsubsection{Listening test}

We conduct an ABX test to compare the style transfer performance between the proposed approach and the baseline approaches. The reference X indicates the source speech. Converted samples (A and B) are presented and the listeners are asked to determine which one has closer speaking style to the reference. The listeners are asked to focus on speaking styles such as tone, stress, speaking speed, phrasing, pausing and etc. 20 testing cases were used. The ABX test results are shown in Figure~\ref{fig:style}. We can see that the proposed approach has significantly better style transfer performance than both baseline approaches.

\section{Conclusions}

In this study, we have presented a source-speaking-style transferable non-parallel VC approach. We incorporate a rhythm module into the sequence-to-sequence VC model, resulting in PER of 5.6\% in the converted speech. Subjective listening tests have also shown its superiority in VC performance in terms of speech naturalness and speaker similarity of the converted speech. We adopt a GST encoder for style inference from the source/reference speech. Experimental results validate the source-style transferability of the proposed approach. This study is an initial attempt to model speaking style for VC. Our future work includes disentangling speaking style, linguistic information and speaker identity from the source speech effectively in an unsupervised way.


\bibliographystyle{IEEEtran}

\bibliography{mybib}


\end{document}